\UseRawInputEncoding
\documentclass[%
 reprint,
 amsmath,amssymb,
 aps,
]{revtex4-2}

\usepackage{graphicx}% Include figure files
\usepackage{dcolumn}% Align table columns on decimal point
\usepackage{bm}% bold math
\begin{document}

\preprint{APS/123-QED}

\title{Threshold current of field-free perpendicular magnetization switching using anomalous spin-orbit torque}% Force line breaks with \\

\author{Tian Yi Zhang$^{1}$, Cai Hua Wan$^{1,\ 2*}$ and Xiu Feng Han$^{1,2,3}$}

 \email{xfhan@iphy.ac.cn; wancaihua@iphy.ac.}
\affiliation{%
 ${}^{1}$Beijing National Laboratory for Condensed Matter Physics, Institute of Physics, University of Chinese Academy of Sciences, Chinese Academy of Sciences, Beijing 100190, China\\
 ${}^{2}$Center of Materials Science and Optoelectronics Engineering, University of Chinese Academy of Sciences, Beijing 100049, China\\
 ${}^{3}$Songshan Lake Materials Laboratory, Dongguan, Guangdong 523808, China
}%

\date{\today}% It is always \today, today,
             %  but any date may be explicitly specified

\begin{abstract}
Spin-orbit torque (SOT) is a promising technique for next-generation magnetic random-access memory (MRAM). Recent experiments have shown that materials with low-symmetry crystalline or magnetic structures can generate anomalous SOT with an out-of-plane component, which is crucial for switching the perpendicular magnetization of adjacent ferromagnetic (FM) layers in a field-free condition. In this study, we derive the threshold current for field-free perpendicular magnetization switching using anomalous SOT and numerically calculate the magnetic moment trajectory in an FM free layer for currents smaller and greater than the threshold current. We also investigate the dependence of switching time and energy consumption on applied current, finding that the minimum energy consumption decreases with an increasing out-of-plane torque proportion. Additionally, we explore the relationships between the threshold current and anisotropy strength, out-of-plane torque proportion, FM free layer thickness, and Gilbert damping constant. The results show a negative correlation between the threshold current and out-of-plane torque proportion, and positive correlations with the other three parameters. Finally, we demonstrate that even when the applied current is smaller than the threshold current, it can still add an effective exchange bias field $H_{{bias}}$ on the FM free layer. The $H_{{bias}}$ is proportional to the applied current $J_{SOT}$, facilitating the determination of anomalous SOT efficiency. Our findings provide insights into the design of spintronic devices that favor field-free switching of perpendicular magnetization using anomalous SOT and offer a means of adjusting the exchange bias field to control FM layer magnetization depinning.
\end{abstract}

%\keywords{Suggested keywords}%Use showkeys class option if keyword
                              %display desired
\maketitle

%\tableofcontents

\section{\label{sec:level1}INTRODUCTION}

The spin-orbit torque (SOT) is a promising technique for developing the next-generation magnetic random-access memory (MRAM) \cite{Kong_2020,Manchon_2019,Mihai_2010,Sinova_2015,Song_2021,Zhang_2018}. Perpendicularly magnetized ferromagnetic (FM) layers have superior performance in thermostability, high density, and retention compared to in-plane magnetized FM layers when used in MRAM \cite{Brataas_2012}. However, ordinary SOT cannot easily switch the perpendicular FM films in the field-free condition. Therefore, determining how to switch the perpendicular magnetization of the FM free layer in the magnetic tunnel junctions (MTJ) by SOTs has long been a frontier of SOT studies. According to the spin Hall effect (SHE) \cite{Sinova_2015}, when electron current ${\bf{{j}}}_{\bf{{c}}}$ is sourced along the x direction, the spin current ${\bf{{j}}}_{\bf{{s}}}$ transports along the z direction and its polarization ${\bf{\sigma }}\propto {\bf{{j}}}_{\bf{{s}}}\times {\bf{{j}}}_{\bf{{c}}}$ will be along the y direction. The adjacent FM free layer is thus affected by the spin current dominatingly through a damping-like SOT ${\bf{\tau }}_{\bf{{d}}}\propto \left({\bf{m}}\times {\bf{\sigma }}\right)\times {\bf{m}}$, which is also along the y direction. This torque is orthogonal to the perpendicular easy axis of the free layer; therefore, using pure SOT alone, we cannot deterministically switch the perpendicular magnetization. Several attempts have been made to circumvent this problem, such as applying an in-plane magnetic field \cite{Liu_2012,Pai_2012}, using structural asymmetry \cite{Chen_2019,Yu_2014}, mediating an in-plane exchange bias/coupling field \cite{Liu_2019,Baek_2018,Fukami_2016,Kong_2019,Lau_2016,Oh_2016}, mediating the interlayer Dzyaloshinskii-Moriya interaction \cite{He_2022} or exploring materials with low-symmetric crystalline or magnetic structures to generate an anomalous SOT \cite{MacNeill_2017,Liu_2019,Bai_2021,Go_2022,Kao_2022,Zhou_2019,Zhou_2020}.

Especially, the groundbreaking studies on low-symmetry materials have shown that the spin polarization, denoted by ${\bf{\sigma }}$, of an out−of−plane−transporting spin current ${\bf{{j}}}_{\bf{{s}}}$ can have both in-plane and out-of-plane components, despite being generated by an in-plane electron current ${\bf{{j}}}_{\bf{{c}}}$. These crystallized materials include Mn${}_{3}$Ir \cite{Liu_2019}, Mn${}_{3}$Pt \cite{Bai_2021}, Mn${}_{3}$Sn \cite{Go_2022}, WTe${}_{2}$ \cite{Kao_2022}, CuPt \cite{Liu_2021}, and more. This anomalous SOT is highly dependent on the crystal or magnetic symmetry. By utilizing the out-of-plane component of ${\bf{\sigma }}$, one can achieve a deterministic switch of the magnetization of a FM free layer without an external magnetic field.

In order to gain a more thorough understanding of the anomalous SOT and its potential for switching a perpendicular magnetization, it is necessary to optimize and utilize relevant parameters that affect the switching dynamics. An analytical derivation of the threshold current in the coexistence of ordinary and anomalous SOTs would be particularly beneficial for this purpose. Despite previous work on formulating threshold currents for different SOT modes \cite{Lee_2013,Sun_2000,Taniguchi_2015,Taniguchi_2016,Zhu_2020}, there is still a lack of a specific threshold current for the coexistence case of the anomalous and ordinary SOTs. This research aims to address this gap.

In this paper, we analytically derive the threshold current required to generate an anomalous SOT for switching the perpendicular magnetization of an adjacent FM layer. Additionally, we use macrospin simulations to investigate the precessional trajectory of the FM layer's magnetic moment both below and above the threshold, finding consistent results with our analytical model. We also examine the dependence of switching time and energy consumption on the applied current, as well as the minimum energy consumption dependence on the proportion of out-of-plane torque. Furthermore, we calculate the threshold current's dependence on anisotropy strength, out-of-plane torque ratio, FM free layer thickness, and Gilbert damping constant. Finally, we demonstrate that an applied current below the threshold can still produce an effective exchange bias field in the FM layer, and provide the relationship between the effective exchange bias field and the applied current. This work can be instructive to design SOT devices with the anomalous SOT materials.

\section{\label{sec:level1}MODEL AND METHOD}

The schematic diagram of a FM free layer magnetization switching driven by the anomalous SOT is shown in Fig. 1. The FM free layer with perpendicular magnetic anisotropy (PMA) is adjacent to a material with low-symmetric crystalline structure. The applied electron current ${\bf{{J}}}_{\bf{{SOT}}}$ is along the -y direction, spin current ${\bf{{J}}}_{\bf{{S}}}$ is along the z direction and the polarization ${\bf{\sigma }}$ of the spin current has components in both x and z directions as shown in the upper left pannel of Fig. 1. $\beta $ is the angle between ${\bf{\sigma }}$ and the x axis. At the interface of the low-symmetric material with the FM free layer, a pure spin current with the ${\bf{\sigma }}$ polarization diffuses into the FM free layer and acts a damping-like SOT on the latter. Then the magnetization of the FM layer will precess around an effective magnetic field or switch its magnetization under the concerted interplay of the SOT with other torques from built-in fields.

\begin{figure}[h]
\includegraphics[width = \linewidth]{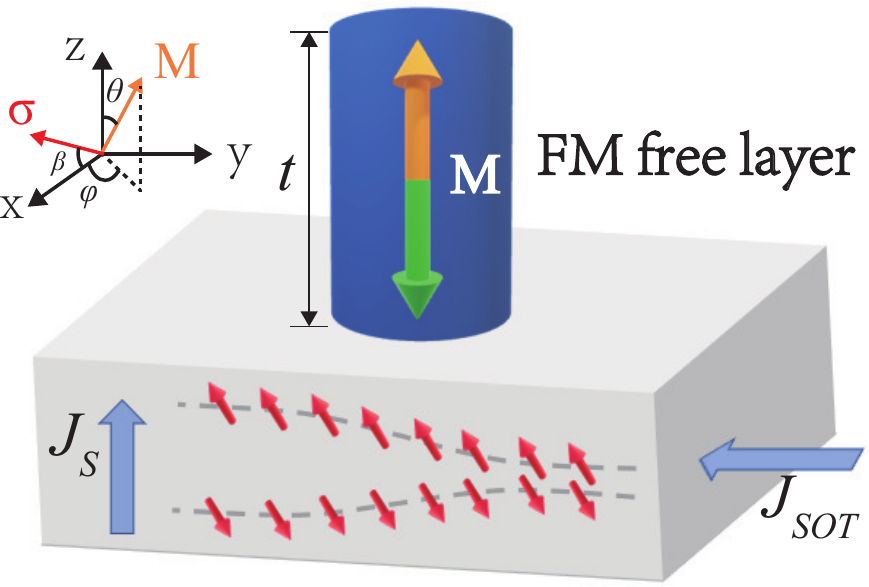}% Here is how to import EPS art
\caption{\label{fig:epsart} A schematic diagram of the FM free layer magnetization switch driven by an anomalous SOT with both in-plane and out-of-plane components. The applied electron current ${\bf{{J}}}_{\bf{{SOT}}}$ is along the -y direction and generates a spin current ${\bf{{J}}}_{\bf{{S}}}$ propagating along the z direction. The spin current diffuses into the FM free layer with the perpendicular magnetic anisotropy to drive its magnetization dynamics.}
\end{figure}

The spin dynamics of the FM layer can be described by the LLG formula \cite{Gilbert_2004}
\begin{equation} \label{GrindEQ__1_}
\begin{array}{ll}
\frac{{\partial }{\bf{m}}}{{\partial }t}=&-\gamma {\mu }_0\left({\bf{m}}\times {\bf{{H}}}_{\bf{{K}}}\right)+\alpha \left({\bf{m}}\times \frac{{\partial }{\bf{m}}}{{\partial }t}\right)\\
&+\gamma {\mu }_0H^{{DL}}_{{SOT}}\left(\left({\bf{m}}\times {\bf{\sigma }}\right)\times {\bf{m}}\right)
\end{array} 
\end{equation} 
where ${\bf{m}}=\frac{{\bf{m}}}{M_s}$ is the unit vector along the direction of magnetization, ${\bf{m}}$ is magnetic moment, $M_s$ is the saturated magnetization value, $\gamma =1.76\times {10}^{11}\ {{T}}^{{-}{1}}{{s}}^{{-}{1}}$ is the gyromagnetic ratio, ${\mu }_0=4\pi \times {10}^{-7}{\ V}{{·}}{s}{{·}}{{A}}^{{-}{1}}{{m}}^{{-}{1}}$ is the permeability of vacuum, ${\bf{{H}}}_{\bf{{K}}}=H_Km_z{\bf{{e}}}_{\bf{{z}}}$ is the anisotropic field, $\alpha $ is Gilbert damping constant, ${\bf{\sigma }}=\left({\sigma }_x,0,{\sigma }_z\right)=\left({cos}\beta ,0,{sin}\beta \right)$ is the unit vector along the electron spin polarization direction, $\beta $ is the angle between the spin polarization direction and the x axis, $H^{DL}_{SOT}$ is the torque intensity generated by SOT, which can be calculated by the following formula \cite{Garello_2013,Khvalkovskiy_2013,Kim_2013,Pai_2014,Park_2014}
\begin{equation} \label{GrindEQ__2_} 
H^{DL}_{SOT}=\frac{J_{SOT}{\theta }_{SH}\hslash }{2et{\mu }_0M_s} 
\end{equation} 
where $J_{SOT}$ is the magnitude of the applied electron current density, ${\theta }_{SH}$ is the spin Hall angle that represents conversion efficiency from electron current to spin current, $\hslash =1.05\times {10}^{-34}{\ J}{\cdot }{s}$ is the reduced Planck constant, $e=1.6\times {10}^{-19}{\ C}$ is the elementary charge carried by an electron, and $t$ is the effective thickness of the free layer after subtracting a dead layer.

When a small current is applied, the anomalous SOT acting on the FM free layer is not large enough to switch the magnetization of the FM layer. The magnetization will precess under the SOT effect, and stabilize to a final direction due to the Gilbert damping. This direction is so-called the direction of the effective field ${\bf{{H}}}_{{eff}}$. Here we constrain ourself in a field-free system which is exactly needed in practice. When the applied current is above a threshold ${\bf{{J}}}_{\bf{{c}}}$, the torque acting on the FM free layer becomes large enough to make the magnetization precession amplitude divergently increase and finally realize magnetization reversal. In the following, we will give the analytical derivation of the threshold current ${\bf{{J}}}_{\bf{{c}}}$. For those readers who interest the dependence of ${\bf{{J}}}_{\bf{{c}}}$ on various material parameters, they can directly skip to Eq. \eqref{GrindEQ__15_} where the final results are directly gave out.

By crossing ${\bf{m}}$\textbf{ }left at both sides of Eq. \eqref{GrindEQ__1_}, we can reform the LLG formula Eq. \eqref{GrindEQ__1_} as in Eq. \eqref{GrindEQ__3_},
\begin{equation} \label{GrindEQ__3_} 
 \begin{array}{cc}
\frac{{\partial }{\bf{m}}}{{\partial }t}= & \frac{-\gamma {\mu }_0}{1+{\alpha }^2}[({\bf{m}}\times {\bf{{H}}}_{\bf{{K}}})+\alpha {\bf{m}}\times ({\bf{m}}\times {\bf{{H}}}_{\bf{{K}}}) \\ 
 & -H^{{DL}}_{{SOT}}(({\bf{m}}\times {\bf{\sigma }}\bf{)}\times {\bf{m}})-\alpha H^{{DL}}_{{SOT}}({\bf{m}}\times {\bf{\sigma }})] 
 \end{array}
\end{equation} 
Let $\frac{{\partial }{\bf{m}}}{{\partial }t}=0$ we can get
\begin{equation} \label{GrindEQ__4_} 
{\bf{m}}{\times }{\bf{{H}}}_{{eff}}{=0} 
\end{equation} 
where the effective magnetic field ${\bf{{H}}}_{{eff}}$ can be written as
\begin{equation} \label{GrindEQ__5_}
\begin{array}{ll}
{\bf{{H}}}_{{eff}}&{=}{\bf{{H}}}_{{K}}{-}H^{DL}_{SOT}\left(\bf{{\sigma }}{\times }{\bf{m}}\right)\\
&{=} H_K(\frac{H^{DL}_{SOT}}{H_K}{sin}\beta m_y{,-}\frac{H^{DL}_{SOT}}{H_K}{sin}\beta m_x{+}\frac{H^{DL}_{SOT}}{H_K}{cos}\beta m_z\\
& \quad {,-}\frac{H^{DL}_{SOT}}{H_K}{cos}\beta m_y{+}m_z) 
\end{array}
\end{equation} 
The direction of ${\bf{{H}}}_{{eff}}$ is also the finally stabilized direction of the magnetization as $J_{SOT}$${}_{\ }$${<}$ $J_c$. From Eqs. \eqref{GrindEQ__4_} and \eqref{GrindEQ__5_}, we can then get

\begin{equation} \label{GrindEQ__6_} 
\left\{ \begin{array}{c}
\frac{H^{DL}_{SOT}}{H_K}{{sin}\beta  m_y\ }{=}km_x \\ 
-\frac{H^{DL}_{SOT}}{H_K}{sin}\beta m_x+\frac{H^{DL}_{SOT}}{H_K}{cos}\beta m_z{=}km_y \\ 
-\frac{H^{DL}_{SOT}}{H_K}{cos}\beta m_y+m_z{=}km_z \end{array}
\right. 
\end{equation} 
where the non-zero real number {k} satisfies
\begin{equation} \label{GrindEQ__7_} 
k^{{3}}\ {-\ }k^{{2}}{+}{\left(\frac{H^{DL}_{SOT}}{H_K}\right)}^2{\ }k\ {-\ }{\left(\frac{H^{DL}_{SOT}}{H_K}\right)}^2{{sin}}^{{2}}\beta {=0} 
\end{equation} 
From Eq. \eqref{GrindEQ__7_}, we get $k=k\left(\frac{H^{DL}_{SOT}}{H_K},\beta \right)$, Then the polar and azimuth angles $\left({\theta }_H,{\varphi }_H\right)\ $of the magnetization in the steady state can be obtained, the schematic diagram of polar angle ${\theta }_H$ and azimuth angle ${\varphi }_H$ in spherical coordinates is shown in the upper left pannel of Fig. 1.
\begin{equation} \label{GrindEQ__8_}
\begin{array}{ll}
{\theta }_H{=arctan}\left(\frac{\left(1-k\right)\sqrt{sin^2\beta +{\left(\frac{kH_K}{H^{DL}_{SOT}}\right)}^2}}{kcos\beta }\right),\\
{\varphi }_H{=arctan}\left(\frac{kH_K}{H^{DL}_{SOT}sin\beta }\right)  
\end{array}
\end{equation} 
After getting ${\theta }_H$ and ${\varphi }_H$, we can transform the coordinate system from the original system O to a new one O' in which ${\bf{{H}}}_{{eff}}$ is directed at the z' axis, and the corresponding transformation matrix between the two coordinates is
\begin{equation} \label{GrindEQ__9_} 
R=\left( \begin{array}{ccc}
{{cos} {\theta }_H\ } & 0 & -{{sin} {\theta }_H\ } \\ 
0 & 1 & 0 \\ 
{{sin} {\theta }_H\ } & 0 & {{cos} {\theta }_H\ } \end{array}
\right)\left( \begin{array}{ccc}
{{cos} {\varphi }_H\ } & {{sin} {\varphi }_H\ } & 0 \\ 
-{{sin} {\varphi }_H\ } & {{cos} {\varphi }_H\ } & 0 \\ 
0 & 0 & 1 \end{array}
\right) 
\end{equation} 
And the relationship from the (x, y, z) coordinate to the (x', y', z') coordinate is simply
\begin{equation} \label{GrindEQ__10_} 
\left( \begin{array}{c}
{{x}}^{{'}} \\ 
{{y}}^{{'}} \\ 
{{z}}^{{'}} \end{array}
\right)=R\left( \begin{array}{c}
{x} \\ 
{y} \\ 
{z} \end{array}
\right) 
\end{equation} 
The transformed coordination allows us to analyze dynamic stability of system straightforwardly. When $J_{{SOT}}{<}J_c$, the components of the magnetization along the x' and y' directions will converge to 0 after a long-enough damping, and as $J_{{SOT}}{\ge }J_c$, the precession amplitude will go divergently and the magnetization will switch to the opposite. At this time, the magnetization along the x 'and y' components will gradually increase, which is our criterion to determine $J_c$. Specifically, considering the two magnetization components along the x' and y' directions, we rewrite the LLG formula Eq. \eqref{GrindEQ__3_} in the following Eq. \eqref{GrindEQ__11_}:
\begin{equation} \label{GrindEQ__11_} 
-\frac{1+{\alpha }^2}{\gamma {\mu }_0}\frac{d}{dt}\left( \begin{array}{c}
m_{x^{{'}}} \\ 
m_{y^{{'}}} \end{array}
\right)={\bf{m}}\left( \begin{array}{c}
m_{x^{{'}}} \\ 
m_{y^{{'}}} \end{array}
\right)+\bf{{G}} 
\end{equation} 
Where ${\bf{m}}$ and $\bf{{G}}$ are $2\ {\times }2$ matrices, and their respective components are explicitly shown below
\begin{widetext}
\begin{equation} \label{GrindEQ__12_} 
\left\{ \begin{array}{ccc}
M_{{11}} & {=} & H^{DL}_{SOT}\left(-{sin\beta cos}^2{\theta }_H-cos{\theta }_Hcos{\varphi }_Hcos\beta sin{\theta }_H\right)+{\alpha (cos}^4{\theta }_H-{cos}^2{\theta }_H{sin}^2{\theta }_H)H_{K{\ }} \\ 
M_{{12}} & {=} & -\alpha H^{DL}_{SOT}\left(sin\beta cos{\theta }_H+cos{\varphi }_Hcos\beta sin{\theta }_H\right)+{cos}^3{\theta }_HH_K \\ 
M_{{21}} & {=} & \alpha H^{DL}_{SOT}\left(sin\beta cos{\theta }_H+cos{\varphi }_Hcos\beta sin{\theta }_H\right)+(cos{\theta }_H{sin}^2{\theta }_H-{cos}^3{\theta }_H)H_K \\ 
M_{{22}} & {=} & H^{DL}_{SOT}\left(-{sin\beta cos}^2{\theta }_H-cos{\theta }_Hcos{\varphi }_Hcos\beta sin{\theta }_H\right)+{\alpha cos}^4{\theta }_HH_{K{\ }} \end{array}
\right. 
\end{equation}
\begin{equation} \label{GrindEQ__13_} 
\left\{ \begin{array}{c}
G_1=-\alpha H^{DL}_{{SOT\ }}cos{\theta }_Hcos\beta sin{\varphi }_H-H^{DL}_{SOT}\left({cos}^3{\theta }_Hcos{\varphi }_Hcos\beta -{sin\beta cos}^2{\theta }_Hsin{\theta }_H\right)+{\alpha cos}^4{\theta }_Hsin{\theta }_HH_{k{\ }} \\ 
{G_2=-\alpha H^{DL}_{SOT}\left({cos}^2{\theta }_Hcos{\varphi }_Hcos\beta -sin\beta cos{\theta }_Hsin{\theta }_H\right)-H^{DL}_{SOT}cos}^2{\theta }_Hcos\beta Sin{\varphi }_H-{cos}^3{\theta }_Hsin{\theta }_HH_{K{\ }} \end{array}
\right. 
\end{equation}
\end{widetext}
From Eq. \eqref{GrindEQ__12_}, we can see that the eigenvalue of the $\ 2\times 2$ matrix ${\bf{m}}$ is ${\lambda }_{1,2}=\frac{M_{11}+M_{22}\pm i\sqrt{-4M_{12}M_{21}-{(M_{11}-M_{22})}^2}}{2}$. When $M_{11}+M_{22}<0$, {m${}_{x}$}${}_{'}$ and {m${}_{y}$}${}_{'}$ decay to 0 if any; in contrast, when $M_{11}+M_{22}>0$, they will diverge once the emergence of an even tiny {\textbar}{m${}_{x}$}${}_{'}${\textbar} or {\textbar}{m${}_{y}$}${}_{'}${\textbar} activated by thermal fluctuations or other reasons. Therefore, the switching criteria turns clear as
\begin{equation} \label{GrindEQ__14_} 
M_{11}+M_{22}{=0} 
\end{equation} 
The threshold current value $J_c$ can be obtained from this condition. Detailed derivation steps are shown in Appendix A, if $\frac{H^{DL}_{SOT}}{H_K}\ll 1$ (widely applicable for most cases), we can get that
\begin{equation} \label{GrindEQ__15_} 
J_c{=}\frac{e{\mu }_0M_sH_Kt}{{\hslash }{\theta }_{SH}}\frac{{4}\alpha }{\sqrt{sin^{{2}}\beta {+16}{\alpha }^{{2}}cos^{{2}}\beta }{+}sin\beta } 
\end{equation} 
Worth noting, according to recent experiment data \cite{Kao_2022}, two typical values of $\frac{H^{DL}_{SOT}}{H_K}$ are $0.014\ and\ 0.023$ at {J}${}_{c}$, so the simplification condition $\frac{H^{DL}_{SOT}}{H_K}\ll 1$ holds reasonable here. When the out-of-plane torque is 0 or $\beta =0$, the result $J_c{=}\frac{e{\mu }_0M_sH_Kt}{{{\hslash }\theta }_{SH}}$ becomes simplified in accordance with the previously proposed{ J${}_{c}$} for the z-type SOT magnetization reversal at a small applied magnetic field \cite{Lee_2013,Zhu_2020,Han_2021}. More interesting, if $\frac{\alpha }{tan\beta }\ll 1$ (a small damping in the order of 10${}^{-2}$ and a substantial anomalous SOT ratio not lower than ${\sim}$0.1 can qualify the condition), the above equation can be simplified as
\begin{equation} \label{GrindEQ__16_} 
J_c=\frac{2e{\mu }_0M_sH_K\alpha t}{\hslash {\theta }_{SH}sin\beta } 
\end{equation} 
This threshold current density then shares a similar fashion with the case of the spin-transfer torque switching mode for the perpendicular MTJ with polarization {P} of the pinned layer replaced by the anomalous spin Hall angle ${\theta }_{SH}sin\beta $.

\section{\label{sec:level1}RESULTS AND DISSCUSSIONS}

We visualize the magnetization trajectory with different out-of-plane torque ratio $\eta\  {\equiv }\ {tan}\beta $ and $J_{{SOT}}$, as shown in Fig. 2. Time step is set as ${d}t\ =\ 1{\ fs}$. The initial direction of ${\bf{m}}$ is along the (0,0,1) in the O coordinate system, Simulation parameters are displayed in TABLE \ref{tab:table1} \cite{Bai_2021,Kim_2018,Lourembam_2018}. For the situation without any out-of-plane SOT or $\eta =0$, when $J_{{SOT}}=1.8\times {10}^{13}{\ A}{{·}}{{m}}^{{-2}}$ which is unable to destabilize the magnetization in the FM free layer, ${\bf{m}}$ is finally stabilized at the direction of the equivalent effective field $\left(0.000,\ -0.451,\ 0.893\right)$. As $J_{{SOT}}=1.9\times {10}^{13}{\ A}{{·}}{{m}}^{{-2}}$, the SOT is large enough to destabilize ${\bf{m}}$\textbf{ }to $\left(-1,\ 0,\ 0\right)$ in the equatorial plane as shown in Fig. 2(a) and (b). These scenarios produce the case of z-type mode without an external bias field. As for $\eta \neq 0$, the final state of $J_{{SOT}}>J_c$ becomes different. From Eq. \eqref{GrindEQ__15_}, we can directly calculate that the threshold currents for $\eta =0.1$ and 0.75 are $J_{{c}}=6.2\times {10}^{12}\ {A}{{·}}{{m}}^{{-2}}$ and $1.1\times {10}^{12}\ {A}{{·}}{{m}}^{{-2}}$ respectively. As $J_{{SOT}}=6\times {10}^{12}{\ A}{{·}}{{m}}^{{-2}}$ and $1\times {10}^{12}\ {A}{{·}}{{m}}^{{-2}}$ for $\eta =0.1$ and 0.75 respectively, the SOT acting on the FM free layer is not large enough, so the precession amplitude gets smaller and smaller, and finally ${\bf{m}}$ is stabilized at the direction of \textbf{H}${}_{eff}$ $\left(0.003,-0.135,\ 0.991\right)$ and $\left(-0.006,-0.0222,\ 0.9997\right)$ for $\eta =0.1$ and 0.75 respectively, see Fig. 2(c) and (e). When $J_{{SOT}}=7\times {10}^{12}{\ A}{{·}}{{m}}^{{-2}}$ and $2\times {10}^{12}\ {A}{{·}}{{m}}^{{-2}}$ for $\eta =0.1$ and 0.75 respectively, the precession amplitude is divergently increasing, and \textbf{m} eventually turns to the opposite direction $\left(-0.0035,-0.1575,\ -0.9875\right)$ and $\left(-0.0003,0.0215,\ -0.9998\right)$ for $\eta =0.1$ and 0.75 respectively, as shown in Fig. 2(d) and (f), consistent with our previous analysis. Worth mentioning, as $J_{{SOT}}>J_{{c}}$, \textbf{m} will not converge to the direction of \textbf{H${}_{eff}$} since the prerequisite for the calculation is ${|m}'_x|,{|m}'_y|\ll 1$, which is violated in this case. This notice does not undermine the strictness of the criteria of deriving $J_{{c}}$. 
\begin{table}[h]%The best place to locate the table environment is directly after its first reference in text
\caption{\label{tab:table1}%
Parameters for numerical calculation (unless otherwise noted).
}
\begin{ruledtabular}
\begin{tabular}{lcc}  
Parameters & Quantity & Value \\
\colrule
{Damping constant} & {$\alpha $} & {0.015 \cite{Lourembam_2018}} \\ 
{Anisotropic field} & {${\mu }_0H_{{K}}$} & {0.85 T \cite{Kim_2018}} \\
{Saturated Magnetization} & {$M_s$} & {${1}{.3}\times {10}^6$ A/m \cite{Kim_2018}} \\
{Ratio of anomalous SOT} & {${tan}\beta $} & {0.1 or 0.75 \cite{Bai_2021}} \\
{the FM thickness} & {$t$} & {1 nm} \\
{Overall spin Hall angle} & {${\theta }_{{SH}}$} & {0.075 \cite{Bai_2021}} \\
\end{tabular}
\end{ruledtabular}
\end{table}
\begin{figure}[h]
\includegraphics[width = \linewidth]{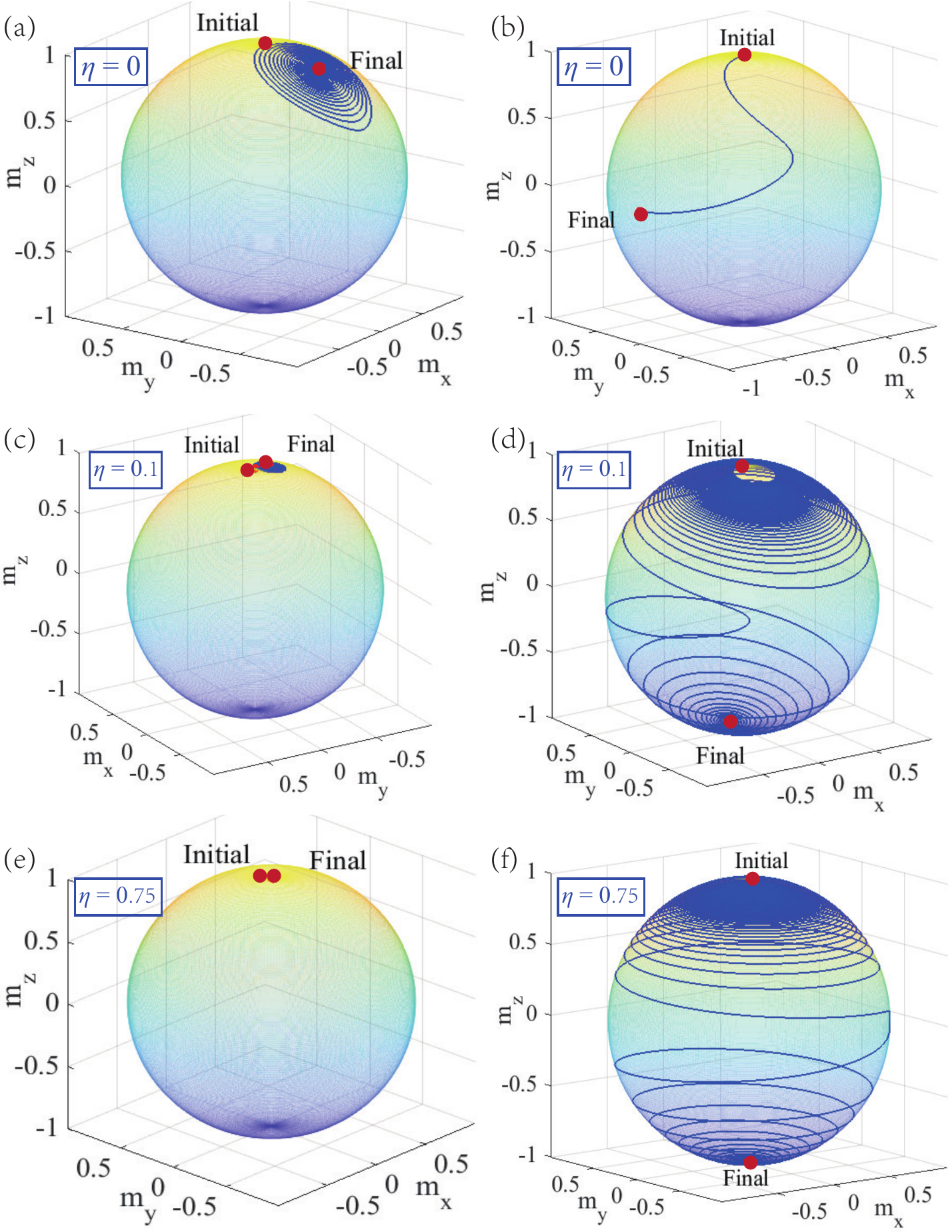}% Here is how to import EPS art
\caption{\label{fig:epsart} The magnetization trajectory with different $\eta $ and $J_{{SOT}}$. The two parameters are shown as follows, $\eta =0$, (a) $J_{{SOT}}=1.8\times {10}^{13}\ {A}{{·}}{{m}}^{{-2}}$ and (b) $J_{{SOT}}=1.9\times {10}^{13}\ {A}{{·}}{{m}}^{{-2}}$; $\eta =0.1$, with threshold current value $J_{{c}}=1.1\times {10}^{12}{\ A}{{{·}}{m}}^{{-2}}$, (c) $J_{{SOT}}=1\times {10}^{12}\ {A}{{·}}{{m}}^{{-2}}$ and (d) $J_{{SOT}}=2\times {10}^{12}\ {A}{{·}}{{m}}^{{-2}}$; $\eta =0.75$, with threshold current value $J_{{c}}=6.2\times {10}^{12}{\ A}{{{·}}{m}}^{{-2}}$, (e) $J_{{SOT}}=6\times {10}^{12}\ {A}{{·}}{{m}}^{{-2}}$ and (f) $J_{{SOT}}=7\times {10}^{12}\ {A}{{·}}{{m}}^{{-2}}$.}
\end{figure}

We also calculate the relationship between the switching time $t_s$ , switching energy consumption $Q_{SOT}$ and $J_{{SOT}}$ when $J_{{SOT}}>J_{{c}}$. Here $\eta =0.75$, $t_s$ is defined as the time from sourcing current to the occurrence of a negative $m_{{z}}$ component in the calculation, switching energy consumption is defined as $Q_{SOT}\equiv J^2_{SOT}{·}t_s$, which scales with the energy consumed in the switching process. We can see from Fig. 3(a) that as $J_{{SOT}}$ increases, {t${}_{s\ }$}decreases rapidly from 16.5 ns when $J_{{SOT}}=1.2\times {10}^{12}{\ A}{{·}}{{m}}^{{-2}}$ to 0.1 ns when $J_{{SOT}}=1.11\times {10}^{13}{\ A}{{·}}{{m}}^{{-2}}$, and the influence of $J_{{SOT}}$ on {t${}_{s}$} is gradually reduced as $J_{{SOT}}$ increases. From Fig. 3(a), we can see that $Q_{SOT}$ minimizes when $J_{{SOT}}$ is near $2.4\times {10}^{12}{\ A}{{·}}{{m}}^{{-2}}$, with minimum value $Q^{min}_{SOT}=7.7\times {10}^{15}{\ }{{A}}^2{{·}}{s}{{·}}{{m}}^{{-4}}$. Then we study the $\eta $-dependence of $Q^{min}_{SOT}$, as shown in Fig. 3(b). As $\eta $ increases, $Q^{min}_{SOT}$ gradually decreases, manifesting the larger out-of-plane torque ratio results in the less energy consumption. With the resistivity of conductive layer $\rho =200\ {\mu }{\Omega }{{·}}{cm}$ \cite{Bai_2021} and the SOT-channel width $l=100{\ nm}$ and length $d=300{\ nm}$, $Q_{SOT}=1\times {10}^{15}\ {{A}}^2{{·}}{s}{{·}}{{m}}^{{-4}}$ corresponds to an energy consumption of 0.06 pJ.

\begin{figure}[h]
\includegraphics[width = \linewidth]{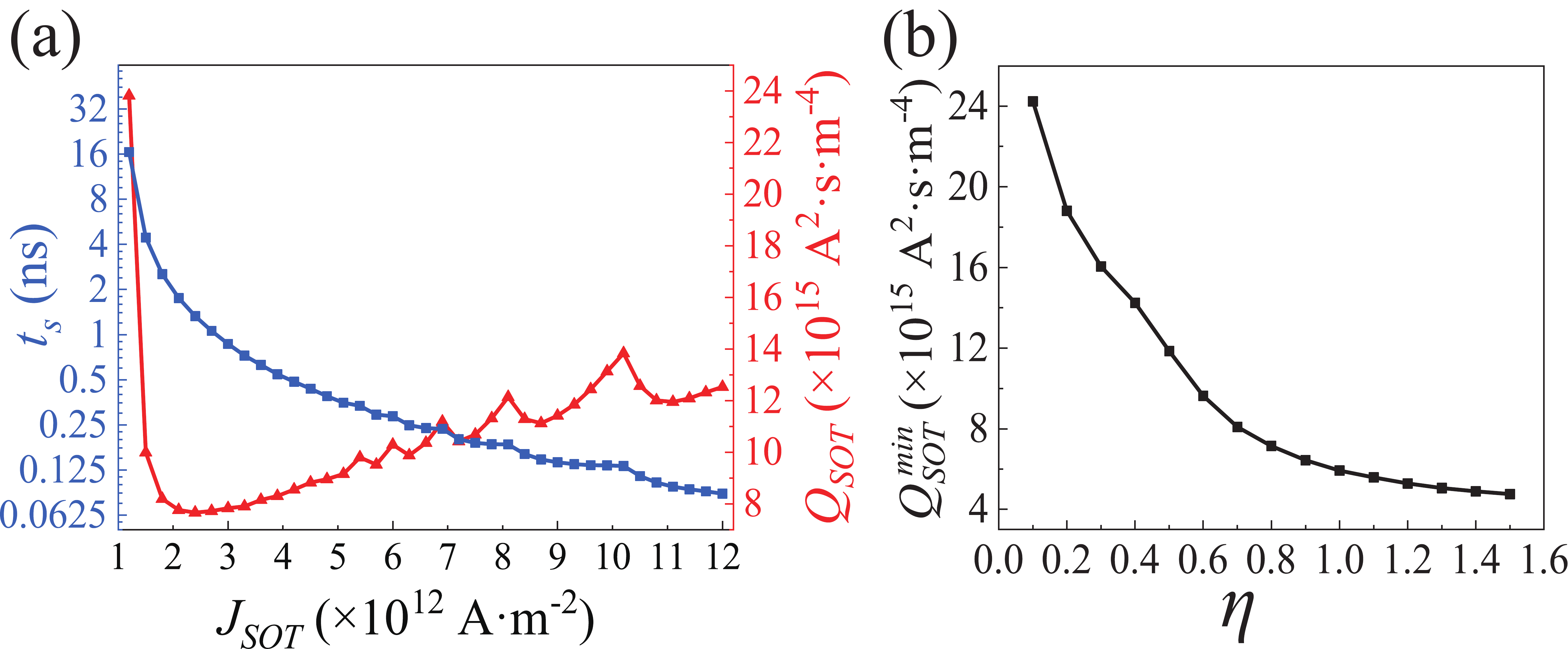}% Here is how to import EPS art
\caption{\label{fig:epsart} The $J_{{SOT}}$-dependence of the switching time {t${}_{s\ }$}and the switching energy consumption $Q_{SOT}$. (b) The $\eta $-dependence of minimum energy loss $Q^{min}_{SOT}$. In the regime of {$\eta$ }${\le}$ 0.8, the increase in {$\eta$} can significantly reduce the value of $Q^{min}_{SOT}$.}
\end{figure}

Then, we numerically calculate the dependence of $J_{{c}}$ on the anisotropy strength $H_{K{\ }}$, out-of-plane torque ratio $\eta $, FM free layer thickness $t$ and Gilbert damping constant $\alpha $, as shown in Fig. 4. We can see that $J_{{c}}$ increases with the increase in {H${}_{K}$} as expected. In addition, we can also see from Fig. 4(a) that $J_{{c}}$ gradually decreases with the increase in {$\eta$}. This is because the decisive factor that affects magnetization switching is the z component of anomalous SOT. When z component of effective field caused by SOT is larger than effective anisotropy field, magnetization switch happens. And as {$\eta$ }increases, the z component of SOT increases, then $J_{{c}}$ becomes lower if we still intend to switch the magnetization. We extract three threshold currents corresponding to different anisotropic properties, as shown in Fig. 4(b). Clearly shown in the figure, the higher anisotropy results in the greater influence of {$\eta$} on $J_{{c}}$.

We also study the thickness $t$ and Gilbert damping $\alpha $ dependences of $J_c$, as shown in Fig. 4(c). As $\alpha $ increases, $J_c$ increases. This feature, similar to the classic spin-transfer torque (STT) switching scheme \cite{Bhatti_2017} , can be explained as following. During the switching dynamics with a larger $\alpha $, \textbf{m} will be easier to converge to its initialized direction due to an enhanced damping which needs a larger $J_c$ to battle against. And as the thickness of the magnetic layer is higher, $J_c$ become larger too with no doubt in accordance with Eq. \eqref{GrindEQ__15_}. We then extract three $J_c$ corresponding to different thicknesses, as shown in Fig. 4(d). $J_c$ scales linearly with $\alpha $, also in accordance with Eq. \eqref{GrindEQ__15_} and the STT scheme. Worth noting, on the other hand, when $\eta =0$, the threshold current Eq. \eqref{GrindEQ__15_} is equal to the threshold current for the z-type SOT mode with an applied magnetic field $H_x$  0. The threshold current for the z-type SOT magnetization switching is\cite{Zhu_2020}
\begin{equation} \label{GrindEQ__16_}
\begin{array}{ll}
J_c&{=}\frac{et{\mu }_0M_sH_K}{{\hslash }{\theta }_{SH}}\\
&[\frac{\sqrt{{4}\alpha {(4}\alpha {+2}\alpha (\frac{H^{FL}_{SOT}}{H^{DL}_{SOT}})^{{2}}{+}\frac{H^{FL}_{SOT}}{H^{DL}_{SOT}}{)+(9}{\alpha }^{{2}}{-}{4}\alpha {(}\frac{H^{FL}_{SOT}}{H^{DL}_{SOT}}{)-8}{\alpha }^{{2}}{(}\frac{H^{FL}_{SOT}}{H^{DL}_{SOT}}{))}{{(}H_x{/}H_K{)}}^{{2}}}}{{{4}\alpha {+2}\alpha (\frac{H^{FL}_{SOT}}{H^{DL}_{SOT}})^{{2}}{+(}\frac{H^{FL}_{SOT}}{H^{DL}_{SOT}}{)}}}\\
&{-}\frac{{5}\alpha {(}H_x{/}H_K{)}}{{4}\alpha {+2}\alpha (\frac{H^{FL}_{SOT}}{H^{DL}_{SOT}})^{{2}}{+(}\frac{H^{FL}_{SOT}}{H^{DL}_{SOT}}{)}}]
\end{array} 
\end{equation} 
with field-like torque intensity $H^{FL}_{SOT}$ and applied magnetic field $H_x$. When the $H^{FL}_{SOT}=0$  and $H_x\ll H_K$, the threshold current is \cite{Lee_2013}
\begin{equation} \label{GrindEQ__17_} 
J_c{=}\frac{et{\mu }_0M_s}{\hslash {\theta }_{SH}}(H_K-\sqrt{2}H_x) 
\end{equation}

\begin{figure}[h]
\includegraphics[width = \linewidth]{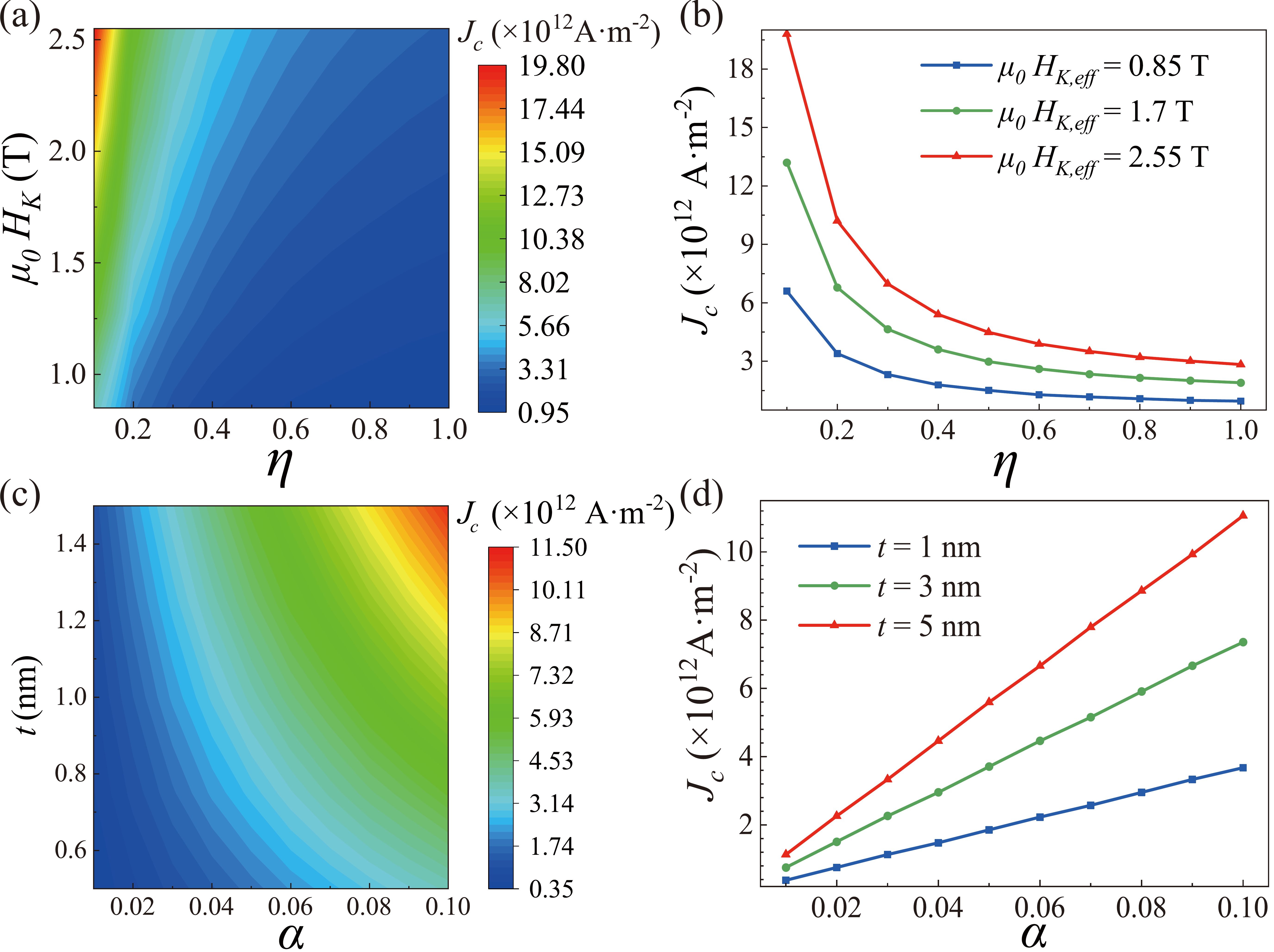}% Here is how to import EPS art
\caption{\label{fig:epsart} The dependence of $J_{{c}}$ on the anisotropic field $H_{K{\ }}$and anomalous ratio $\eta $. (b) The {$\eta$} dependence of $J_{{c}}$ under $\alpha =0.015,\ \ t=1\ {nm}$ extracted from Fig. 4(a). (c) The dependence of $J_{{c}}$on the thickness $t$ and Gilbert damping $\alpha $. (d) The $\eta $ dependence of $J_{{c}}$ under ${\mu }_0H_K=0.85\ {T},\ \eta =0.75$ extracted from Fig. 4(c).}
\end{figure}

The experiment \cite{Liu_2021} and the above derivation have shown that the anomalous SOT can switch the magnetization of FM free layer if the applied current is above the threshold $J_c$. However, even if $J_{SOT}$${}_{\ }$${<}$ $J_c$, the anomalous SOT can still manifest itself by acting an effective exchange bias field $H_{bias}$, which facilitates us to determine the anomalous SOT efficiency. We calculate the hysteresis loops corresponding to different $J_{SOT}$ ${<}$ $J_c$, as shown in Fig. 5(a). Here the LLG equation has to take magnetic field ${\bf{{H}}}_z$ into account.
\begin{equation} \label{GrindEQ__16_}
\begin{array}{ll}
\frac{{\partial }{\bf{m}}}{{\partial }t}=&-\gamma {\mu }_0\left({\bf{m}}{\times }\left({\bf{{H}}}_{\bf{{K}}}{+}{\bf{{H}}}_{\bf{{z}}}\right)\right){+}\alpha \left({\bf{m}}{\times }\frac{{\partial }{\bf{m}}}{{\partial }t}\right)\\
&{+}\gamma {\mu }_0H^{{DL}}_{{SOT}}\left(\left({\bf{m}}{\times }{\bf{\sigma }}\right){\times }{\bf{m}}\right)
\end{array} 
\end{equation} 
Fig. 5(a) shows for an unbiased loop without $J_{SOT}$, the forward and backward switching coercivity of the FM layer is symmetric. However, when a positive (negative) $J_{SOT,0} = 6.28\times10^{11}$ Am$^{-2}$ is applied, the hysteresis loop is biased leftward (rightward) or their H${}_{z}$-symmetrical axis is offset in the negative (positive) direction. Therefore, equivalently, a nonzero $J_{SOT}$ imposes an exchange bias field $H_{bias}$ along the z axis to the FM free layer, which determines the switching direction in the field-free condition. Compared to z-type SOT \cite{Han_2021} that switch magnetization with the help of in-plane magnetic field, this result give a novel method to control magnetization of FM layer, which is easier to integrate. We also calculate the dependence of $H_{{bias}}\ {on}\ {\ J}_{{SOT}}$ shown in Fig. 5(b) which appears linear with each other and the slope is $\frac{{\theta }_{{SH}}\hslash {sin}\beta }{2et{\mu }_0M_s\alpha }=6.05\times {10}^{-7}\ m,$ if $\frac{H^{DL}_{SOT}}{H_K}\ll 1$ and $\frac{H_{bias}}{H_K}\ll 1$. This result provides a direct way to determine the anomalous torque efficiency. Compared with previous work that detect anomalous torque with in-plane magnetization of CoFeB \cite{Baek_2018}, our results shows a simpler linear relationship between $H_{bias}$ and ${\ J}_{{SOT}}$.

\begin{figure}[h]
\includegraphics[width = \linewidth]{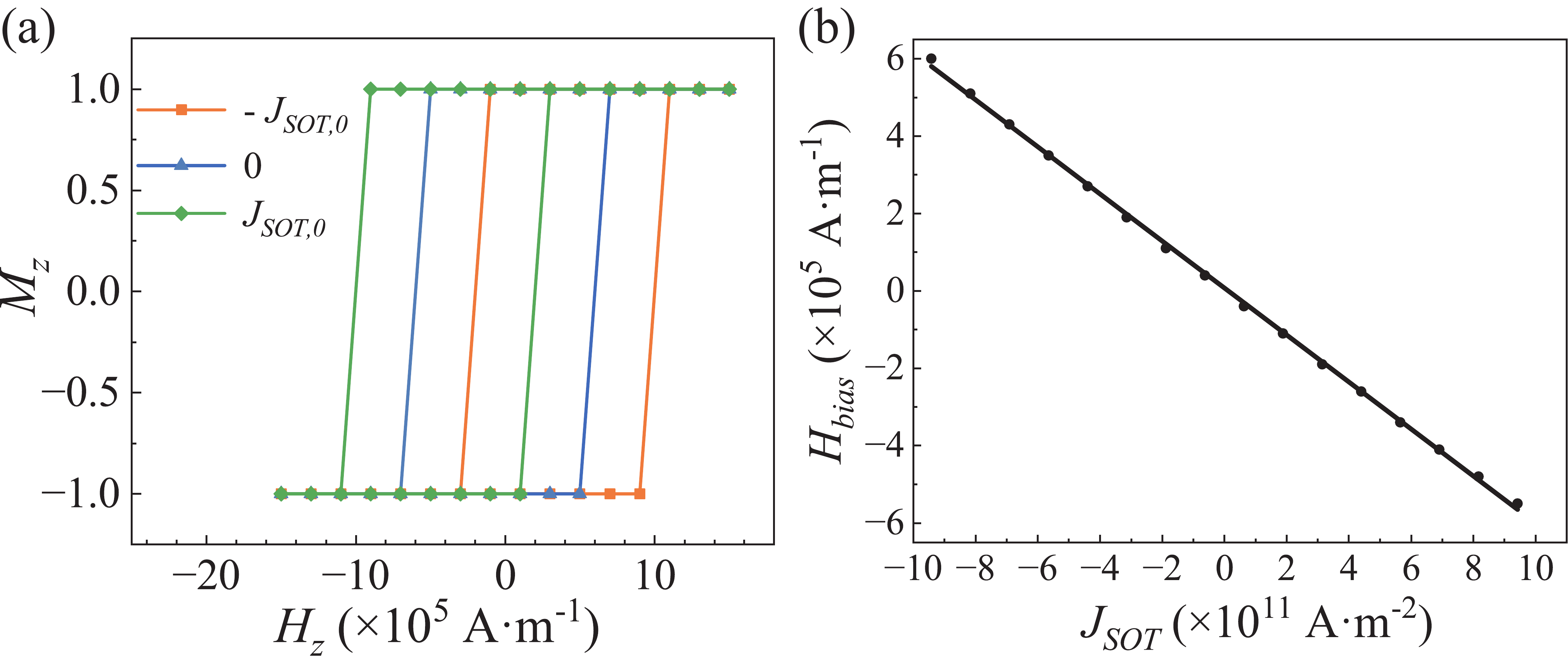}% Here is how to import EPS art
\caption{\label{fig:epsart} (a) Hysteresis loop of FM free layer under different applied currents${\ J}_{{SOT}}$, (b) Relationship between equivalent exchange bias field $H_{bias}$ and applied current${\ J}_{{SOT}}$. Here $\eta =0.75$.}
\end{figure}

\section{\label{sec:level1}CONCLUSIONS}

In this paper, we present an analytical derivation of the threshold current required to achieve field-free switching of perpendicular magnetization using the anomalous SOT in combination with an ordinary one. We also conduct numerical simulations to investigate the magnetization trajectory of a FM free layer when the applied current is both below and above the threshold. Our analytical and numerical results are in agreement. Furthermore, we explore the dependence of the switching time and energy consumption on the applied current and show that the minimum energy consumption is negatively correlated with the out-of-plane torque proportion. Additionally, we investigate the effects of various parameters, including anisotropy strength, out-of-plane torque ratio, FM free layer thickness, and Gilbert damping constant, on the threshold current. Our findings indicate a negative correlation between the out-of-plane torque proportion and the threshold current, and a positive correlation between the other three parameters and the threshold current. Finally, we demonstrate that when the applied current is below the threshold, it can induce an exchange bias field $H_{bias}$ in the FM free layer. Our numerical results show that the exchange bias field $H_{bias}$ is proportional to the applied current $J_{\mathrm{SOT}}$. This study provides insights into the design of spintronic devices that enable field-free switching of perpendicular magnetization using the anomalous spin-orbit torque and offers a direct method for adjusting the exchange bias field, which can be useful in controlling FM layer magnetization pinning and depinning.

\begin{acknowledgments}
This work is financial supported by the National Key Research and Development Program of China (MOST) (Grant No. 2017YFA0206200, 2021YFB3601300), the National Natural Science Foundation of China (NSFC) (Grant No. 12134017 ,11974398 , 12061131012), and partially supported by the Strategic Priority Research Program (B) (Grant No. XDB33000000, Youth Innovation Promotion Association of CAS (2020008)).
\end{acknowledgments}

\appendix
\section{Derivation of Eq. (15)}
\renewcommand{\theequation}{S\arabic{equation}}
From Eq. \eqref{GrindEQ__7_}, suppose $k_1=k-\frac{1}{3}$,  we can get that
\begin{equation}
 k^{\mathrm{3}}_{\mathrm{1}}\mathrm{+}\left({\left(\frac{H^{DL}_{SOT}}{H_K}\right)}^{\mathrm{2}}\mathrm{-}\frac{\mathrm{1}}{\mathrm{3}}\right)k_{\mathrm{1}}\mathrm{+}\left(\mathrm{-}\frac{\mathrm{2}}{\mathrm{27}}\mathrm{+}\left(\frac{\mathrm{1}}{\mathrm{3}}\mathrm{-}sin^{\mathrm{2}}\beta \right){\left(\frac{H^{DL}_{SOT}}{H_K}\right)}^{\mathrm{2}}\right)\mathrm{=0}   
\end{equation}
If $\frac{H^{DL}_{SOT}}{H_K}\ll 1$, we can get that
\begin{equation}
 \begin{array}{ll}
k_1 &= \sqrt[3]{\frac{1}{27}-\frac{1-3sin^2\beta }{6}{\left(\frac{H^{DL}_{SOT}}{H_K}\right)}^2+\frac{sin\beta }{3\sqrt{3}}\frac{H^{DL}_{SOT}}{H_K}}\\
&\quad+\sqrt[3]{\frac{1}{27}-\frac{1-3sin^2\beta }{6}{\left(\frac{H^{DL}_{SOT}}{H_K}\right)}^2-\frac{sin\beta }{3\sqrt{3}}\frac{H^{DL}_{SOT}}{H_K}}\\
&=\frac{2}{3}-cos^2\beta {\left(\frac{H^{DL}_{SOT}}{H_K}\right)}^2
 \end{array} 
\end{equation}
So that
\begin{equation}
k=1-cos^2\beta {\left(\frac{H^{DL}_{SOT}}{H_K}\right)}^2
\end{equation}
We can get the pole angle and azimuth angle $\left({\theta }_H,{\varphi }_H\right)$
\begin{equation}
\left\{ \begin{array}{ll}
tan{\theta }_H&=\frac{\left(1-k\right)\sqrt{sin^2\beta +(kH_K/H^{DL}_{SOT})^2}}{kcos\beta }\\
&=cos\beta \frac{H^{DL}_{SOT}}{H_K} \\ 
cos{\varphi }_H&=\frac{sin\beta H^{DL}_{SOT}/H_K}{\sqrt{(1-(cos\beta H_K/H^{DL}_{SOT})^2+(sin\beta H_K/H^{DL}_{SOT})^2}}\\
&=sin\beta \frac{H^{DL}_{SOT}}{H_K}
\end{array}
\right.  
\end{equation}
From Eq. \eqref{GrindEQ__12_}, we can get threshold current using equation $M_{\mathrm{11}}+M_{\mathrm{22}}=0$:
\begin{equation}
J_c\mathrm{=}\frac{et{\mu }_0M_sH_K}{{\theta }_{SH}\mathrm{\hslash }}\frac{\mathrm{4}\alpha }{\sqrt{sin^{\mathrm{2}}\beta \mathrm{+16}{\alpha }^{\mathrm{2}}cos^{\mathrm{2}}\beta }\mathrm{+}sin\beta } 
\end{equation}

\nocite{*}

% \bibliography{apssamp}% Produces the bibliography via BibTeX.
%apsrev4-2.bst 2019-01-14 (MD) hand-edited version of apsrev4-1.bst
%Control: key (0)
%Control: author (8) initials jnrlst
%Control: editor formatted (1) identically to author
%Control: production of article title (0) allowed
%Control: page (0) single
%Control: year (1) truncated
%Control: production of eprint (0) enabled
\providecommand{\noopsort}[1]{}\providecommand{\singleletter}[1]{#1}%

\end{document}